\documentstyle[preprint,aps,amssymb,epsf]{revtex}
\newcommand{\beq}{\begin{equation}}
\newcommand{\eeq}{\end{equation}}
\newcommand{\bea}{\begin{eqnarray}}
\newcommand{\eea}{\end{eqnarray}}
\begin{document}

\title{Axial phase of quantum fluids in nanotubes}

\author{S. M. Gatica$^{1,2}$, G. Stan$^{2,3}$, M. M. Calbi$^1$, J. K. Johnson$^4$
and M. W. Cole$^{2,5}$}

\address{$^1$Departamento de F$\acute{\imath}$sica,
Universidad de Buenos Aires, Buenos Aires 1428, Argentina.\\
$^2$Department of Physics and Center for Materials Physics,\\
Pennsylvania State University, University Park, PA 16802, USA.\\
$^3$Present address: Institute for Physical Science and Technology and\\ Department of Chemical Engineering,
University of Maryland, College Park, MD 20742, USA.\\
$^4$Department of Chemical and Petroleum Engineering,\\ University of 
Pittsburgh, Pittsburgh, PA, 15260, USA.\\
$^5${corresponding author, e-mail: mwc@psu.edu, phone: 814-863-0165, fax: 814-865-3604}}

\date{\today}
\maketitle
\baselineskip=22pt

\begin{abstract}

We explore the equations of state and other properties of various quantum
fluids ($^3$He, $^4$He, their mixtures, and H$_2$) confined within individual 
carbon nanotubes. Above a threshold number of particles, 
$N_a$, the fluid density near the axis begins to grow above a negligibly small 
value. The properties of this axial fluid phase are sensitive to the tube size 
and hence to the transverse compression in the case of a bundle of nanotubes. 
We consider He at zero temperature and H$_2$ at low temperatures. 
\end{abstract}

\section{Introduction}

The  physical properties of carbon nanotubes are expected to manifest a
variety of intriguing one-dimensional (1d) and quasi-1d behaviors 
\cite{ebbesen,inoue,white,radhakrishnan,cole,steele,stan1,stan2,vidales,peterson,wang1,wang2,stan3}. We, and other groups, have found that fluids 
absorbed within these tubes ought to exhibit an ``effective'' dimensionality 
which is sensitive to the geometry, species, number ($N$) of particles and 
temperature ($T$). For example, in the case of isolated nanotubes, the ideal 
gas regime of very low $N$, is predicted to show crossover between 1d and 2d 
behavior as $T$ increases, due to the excitation of azimuthal motion of the 
atoms adsorbed on the inner wall of the tubes \cite{stan3}. For a bundle of 
nanotubes, in contrast, it has been predicted and found experimentally that 
low $N$ gases which have small diameter are adsorbed strongly in interstitial channels, between tubes, 
in which the gases display 1d motion \cite{cole,stan2,wang1,teizer}. Larger diameter gases adsorb preferentially within the tubes. 
\cite{kuznetsova,uptake}

 The novel collective properties of the absorbed interacting
fluid  at higher $N$ are now being explored. In this paper we consider the evolution, as a
function of $N$, of the structure and properties of dense quantum fluids
within individual nanotubes at $T=0$. Consider He for specificity. Above the
ideal gas regime of density, there appears a regime of density in which
liquid $^4$He is  adsorbed on the walls of the tube; we call this a cylindrical
``shell'' phase \cite{ideal_gas}. At somewhat higher density this fluid should 
solidify; we expect that this film is similar to the much studied 
incommensurate monolayer solid film on graphite \cite{bruch,dash,curvature}. 
For values of N below a threshold value $N_a$ there exists only this shell 
phase. We predict here a transition in which there appears, for $N > N_a$, 
an axial phase of this fluid, signified by the presence of a fluid confined to
 the vicinity of the tube axis. This axial phase transition is qualitatively 
similar to two other transitions familiar in the field of adsorption: 
capillary condensation (CC) and the layering transition \cite{cc,griffiths}. 
While the latter (2d) transition is known to occur at finite $T$, the 
appearance of the axial phase is a genuine thermodynamic transition only at 
$T=0$ because the geometry is 1d, so that fluctuations eliminate the 
transitions at higher $T$. This feature is also the case for a CC transition in
 a single pore, but often in experiments the effect of interactions with other 
pores is such as to permit a CC transition at nonzero $T$ 
\cite{radhakrishnan,cole}.
 Similar behavior of the axial phase transition should occur in a nanotube 
bundle (``rope''), but we have not yet investigated that problem. Also, the 
same phenomenon is expected to occur for a classical system since nothing 
specifically related to quantum effects plays a paramount 
role \cite{steele,wang4}.

We have investigated the axial phase with three distinct methods. The next
section presents Hartree model calculations for four systems: $^4$He, $^3$He,
their mixtures, and H$_2$. These calculations rely extensively on the use of
other workers' calculations and data for 2d quantum systems \cite{whitlock,ni,elgin}.
 Section  III presents  results of 
Path Integral Monte Carlo studies of H$_2$.
 Section \ref{sec:mixture} reports results  for $^3$He-$^4$He 
mixtures, obtained with the Hartree approach. Section \ref{sec:DF} describes 
the results obtained with Density Functional calculations applied to $^4$He 
\cite{gatica}. The comparison between our various approaches is interesting, 
as is that between the present data and ``exact'' results for $^4$He in 1d 
\cite{stan2}. Section \ref{sec:conclusions} summarizes our results and 
conclusions.

\section{Hartree Model Calculations for Helium and Hydrogen}
\label{sec:hartree}

Our goal is to determine the threshold coverage for formation of the axial
phase. The general approach described in this section is analogous to
that used in previous studies of the problem of layer promotion \cite{whitlock,cheng}. The assumption is made that, above threshold, atoms (or molecules 
in the case of H$_2$) form two coexisting phases, the axial phase and the shell
phase. These are characterized by 1d densities $N_a/L$ and $N_s/L$, 
respectively. The condition for equilibrium is that the chemical potentials of 
the two phases coincide:
\beq
\mu_{axial} = \mu_{shell}
\label{eq:mu}   
\eeq
The axial phase is thought of as a 1d phase affected by the ``external''
Hartree potential provided by the shell phase and the host nanotube. Its
chemical potential is assumed to satisfy
\beq
\mu_{axial}(N_a/L) = \epsilon_a + \mu_{1d}(N_a/L)  
\eeq
The 1d approximation ought to be valid here in that we focus on the regime
when the rms displacement transverse to the axis is small compared to the
axial phase interparticle spacing and the axial and shell atoms are well
separated spatially. Here $\mu_{1d}(N_a/L)$ is the chemical potential of a 
1d fluid at density N$_a$/L and $\epsilon_a$ is the eigenvalue of the atoms 
in the Hartree potential. At the threshold for forming the axial phase, 
$\mu_{1d}$ assumes its lowest value, the ground state cohesive energy of the 1d
fluid. This energy has been found previously for $^4$He to be of order 2 mK,
which is negligible small in the present context, so we ignore it for $^4$He
and the other cases studied here \cite{energy}. 

The energy $\epsilon_a$ was computed from a numerical solution of the 
Schr\"odinger equation for ground state (zero angular momentum) atomic motion 
in the Hartree potential:
\beq
V_{tot} (r) = V_C (r) + \theta\ V_H (r)
\label{eq:V_tot}   
\eeq
where $V_C (r)$ is the potential near the axis due to the carbon atoms alone
 computed in the assumption of smooth tube walls as in previous studies 
\cite{stan1}. The second term is a Hartree interaction, where $\theta$ is the 
density of gas atoms/molecules in the shell, and  $\theta\ V_H (r)$ is the
interaction due to the shell:
\beq
V_H (r) = \int d{\bf r'}\ {|\Psi_s ({\bf r'})|^2\ 
U_{gg}(|{\bf r} -{\bf r'}|) }
\label{eq:V_H}  
\eeq
where $\Psi_s ({\bf r'})$ is the wave function of gas 
atoms/molecules in the shell and $U_{gg}$ is the gas-gas interaction. 
We make the further approximation of assuming that the atoms in the shell
 phase are narrowly confined in the radial direction:
\beq
|\Psi_s ({\bf r})|^2 \sim \delta (r-R)
\eeq
where $R$ is the radius of the cylindrical surface near which the atoms are 
situated. Then the evaluation of $V_H$ reduces to integrations over the 
longitudinal and azimuthal coordinates, as shown in Ref. \cite{stan1}.

The ground state eigenfunction of the Schr\"{o}dinger equation for an atom 
close to the axis is of the form $\Psi_a ({\bf r}) = f (r) \ exp(ikz)$, 
where $r$ and $z$ are the cylindrical coordinates of the atom's position 
vector ${\bf r}$. The eigenvalue $\epsilon_a$ is then determined from the 
differential equation satisfied by the ground state radial wave function, 
$f (r)$: 
\beq
\frac{d^2 f}{dr^2} + \frac{1}{r} \frac{df}{dr} + 
\frac{2m}{\hbar^2}\left[\epsilon_a - V_{tot} (r)\right] f(r) = 0
\eeq
We solve this equation numerically.

Figures \ref{fig:mu4} and \ref{fig:mu3} display for the He isotopes the 
two chemical potentials appearing in Eq. \ref{eq:mu}; their curves' crossing 
occurs at the threshold at which the axial phase appears. The abscissa in 
these figures is a 2d density of the shell phase:
\beq
\theta = N_s/ (2 \pi R L)   
\eeq
As seen in figures \ref{fig:mu4} and \ref{fig:mu3}, 
the eigenvalue $\epsilon_a$ is a nearly linear function of the shell phase 
density in the neighborhood of the transition, as expected from a perturbation 
theory of the Hartree interaction, based on Eq. \ref{eq:V_tot}.

We have made the simplest plausible assumption for the function $\mu_{shell}
(\theta)$  - that it coincides with the 2d chemical potential on graphite,
apart from an additive constant arising from the stronger substrate
attraction and resulting larger binding energy in the tubes:
\beq
\mu_{shell} (\theta) = \mu_{graphite} (\theta) + \Delta E_b   
\eeq
Here $\Delta E_b$ is the difference in binding energy between the result for 
the nanotubes \cite{stan3} and that for graphite \cite{elgin,cole2}. The 
justification for this implicit neglect of the effect of curvature appears 
in the Appendix. Our values of these functions $\mu_{shell} (\theta)$ are 
taken from theoretical studies of Ref. \cite{whitlock,ni} which are consistent 
with experiments reported in Ref. \cite{elgin}. 

The key properties of the shell and axial phases at threshold are displayed
in Tables \ref{table:4He} and \ref{table:3He} and Fig. \ref{fig:rho} for 
the isotopes $^4$He and $^3$He, respectively. There are no significant 
differences between the qualitative behaviors of the two isotopes. 
Consider, for example, the case of $^4$He within nanotube having a radius of
$R_C= 6$ \AA. Note that the He shell radius, $R= 3.06$ \AA, is about 3 \AA\ 
closer to the axis than is the carbon shell; $R$ increases by ca. 1 \AA\ as 
$R_C$ increases by 1 \AA. As $R_C$ increases, the energy levels
 of the shell ($E_0$) and axial states at threshold increase monotonically. 
The threshold density $\theta_c$ of the shell state is insensitive to this 
change, remaining close to the known value 0.115 \AA$^{-2}$ for monolayer 
completion density on graphite \cite{ni,bretz}. However, the threshold 
chemical potential $\mu_c$ varies much more; it increases by about 20 K for 
each 1 \AA\ change in $R$; for $R_C$ = 8 \AA, $\mu_c$ approaches 
the value -35 K for monolayer completion on graphite \cite{monolayer}. Note 
that the potential energy responsible for the axial state changes by an even 
larger amount than $\mu_c$. The minimum value of the total potential energy, 
$V_{tot} (r_{min})$, changes by 90 K as $R_C$ increases by 2 \AA, while 
$\mu_c$ changes by only 40 K. The reason is that the zero point kinetic energy 
also changes significantly, as can be understood from the uncertainty 
principle. The kinetic energy is much larger at $R_C = 6$ \AA, for which 
state the root mean square radial coordinate is $r_{rms} = 0.41$ \AA, than at 
$R_C = 8$ \AA, for which $r_{rms} = 2$ \AA. Note from the table that the carbon
 shell's contribution to the potential energy $V_C (r)$ is typically 40\% of 
the total potential energy while the He shell contribution is about 60\%; 
these proportions change slowly as $R_C$ changes. Finally, we note the 
qualitative change in the axial state wave function seen in Fig. \ref{fig:rho}.
 For the case of $R_C = 8$ \AA, the axial state's probability density is no 
longer confined to the immediate vicinity of the axis, exhibiting a maximum 
near $r = 2$ \AA. One observable property which is determined by this spread
is the momentum distribution function, which ought to be much narrower in
the case of such a dispersed wave function. We intend to consider this
topic in future work.

Figures \ref{fig:mu_H2} and \ref{fig:rho_H2} and Table \ref{table:H2} present 
the results of analogous calculations for H$_2$ adsorption in nanotubes of 
varying size. The data corresponding to $R_C = 8$ \AA\ look qualitatively 
similar to those for He except with different numerical values, of course. As 
for $^4$He, the threshold values of coverage and chemical potential are 
similar to those for monolayer completion on graphite 
($\theta_c = 0.094$ \AA$^{-2}$, $\mu_c = -244$ K) \cite{nielsen}. 

For $R_C = 7$ \AA, the axial phase threshold value of the chemical potential 
for H$_2$, i.e. the crossover seen in Fig. \ref{fig:mu_H2}, moves to a much 
lower value ($\mu_c = -385$ K) than for $R_C = 8$ \AA\ (-261 K). For the 
smallest case studied, $R_C = 6$ \AA, we find no crossing of the chemical 
potential curves. This means simply that nanotubes which are so small do not 
produce an axial phase, according to our calculations.

\section{Path Integral Simulations for Hydrogen}
\label{sec:path}

As a check on the validity of 
the Hartree calculations we have performed 
molecular simulations of hydrogen adsorption in nanotubes,  
using the path integral formalism \cite{feynman} 
implemented in the grand canonical ensemble. The algorithm 
we have used follows the 
previous work of Wang and Johnson \cite{wang1998,wang1999}. 
The Silvera-Goldman \cite{silvera} isotropic 
interaction potential was used for the H$_2$-H$_2$ 
potential. The nanotube-hydrogen potential used in the 
path integral simulations was the same as that used in the 
Hartree calculations. 
The path integral calculations were carried out at
finite temperature, 
so we do not expect complete agreement with the Hartree method. 
In particular, the shell-axial first order phase change will disappear 
in favor of a continuous transition as the axial phase is populated. 
Calculations were performed at 5 and 10 K for nanotubes of 
radius 5, 6, 7, and 8 \AA. 
In the path integral formalism each quantum molecule is replaced 
by a classical ring polymer containing some number of beads. 
The accuracy of the simulations depends on the 
number of beads used in each ring, with the results becoming 
exact as the number of beads per ring approaches 
infinity. We found that 50 beads per ring were sufficient to obtain 
 convergent 
results at both 5 and 10 K. 

The shell and axial phase densities of hydrogen in the 7 \AA\ tube are 
plotted in figure~\ref{fig:7Acoverage} as a function of the chemical 
potential. The shell phase density remains fairly constant while the 
axial phase increases dramatically at a chemical potential of about $-375$ K, 
in reasonable agreement with the Hartree value of $\mu_c = -385$ K 
given in Table~\ref{table:H2}. 
The results at a temperature of 5 K (not shown) 
are very similar, but the population of 
the axial phase commences at a somewhat higher chemical potential.  
The shell and axial phase densities for the 
6 \AA\ radius nanotube are shown in figure~\ref{fig:6Acoverage}. The 
path integral calculations demonstrate that population of the 
axial phase does occur, 
with the onset at a chemical potential of about $-325$ K. 
This is inconsistent with the Hartree calculations, which do not show 
the formation of an axial phase at any chemical potential for this small value 
of R. 

Figure~\ref{fig:6Arho} is a plot of 
the density profile in the 6 \AA\ nanotube at three different 
chemical potentials. Note from figure~\ref{fig:6Arho} 
that the location of the shell phase 
is pushed progressively outward in response to the increased population 
of the axial phase. 
The disagreement between the Hartree and path integral calculations 
can be ascribed to the fact that the Hartree model does not account for 
the shift in the location of the shell phase due to the presence of 
the axial phase, and therefore the Hartree model misses the axial 
phase observed in the simulations. The density profile 
for the 7 \AA\ tube is plotted in figure~\ref{fig:7Arho} for a series 
of chemical potentials.  Note that that shell phase is essentially 
unperturbed by the presence of the axial phase in the 7 \AA\ nanotube.

The coverage and density profiles for the 8 \AA\ nanotube are plotted 
in figures~\ref{fig:8Acoverage} and \ref{fig:8Arho}, respectively. Note 
that the onset of axial phase population occurs at a higher chemical 
potential (larger bulk pressure) than for either the 6 or 7 \AA\ nanotube. 
This is somewhat surprising given that the shell phase must be compressed 
in order to make way for the axial phase in the 6 \AA\ tube. 
The onset of population of the axial phase in the 8 \AA\ nanotube occurs 
at about $\mu = -298$ K, in fair agreement with the Hartree value of 
$\mu_c = -261$ K from Table~\ref{table:H2}. 
Note from figure~\ref{fig:8Arho} that it is 
evident that the 8 \AA\ nanotube does 
not actually have an axial phase, but rather a second shell phase, centered 
about 2 \AA\ from the center of the tube. This agrees with the 
Hartree calculations of figure~\ref{fig:rho_H2}.

\section{Helium Mixtures}
\label{sec:mixture}

We now consider the case of isotopic mixtures of helium adsorbed at medium
to high density within the tube. In the analogous situation on the graphite
basal plane, there is complete isotopic separation at $T=0$ \cite{miller}. We 
expect the same behavior in the nanotubes at coverages below that required for producing 
the axial phase. In this case, the criterion determining the distribution of 
the two phases on the surface is equality of their spreading pressures. Figure 
\ref{fig:dens34} presents the coexisting 2d densities of these phases, as 
derived from application of this criterion to the equation of state data of Bruch \cite{bruch2} and Figure \ref{fig:mixture} shows a schematic view of the mixture configuration, determined by computing the threshold for the axial phase.
We found that $^3$He atoms first go to the axial region surrounded by the $^4$He atoms at virtually
the same spreading pressure at which $^4$He atoms begin to go on the axis. The preferential binding in the $^4$He region is a consequence
 of the higher 2d shell density at $^3$He-$^4$He coexistence.  Our numerical results for this
 onset condition  appear in Figure \ref{fig:mu34} and Table \ref{table:mixture}.

\section{Density Functional Study of $^4$He}
\label{sec:DF}

We consider that the energy of an ensemble of $N$ helium atoms in the
nanotube has the form
\beq
E [\rho]=E_0 [\rho]+ \int d{\bf r}\ \rho(r)\ V_C\ (r) 
\eeq
where $V_C (r)$ is the adsorption potential due to all the carbon
 atoms in the nanotube evaluated at point {\bf r} and  $E_0 [\rho]$ is the
Finite Range Density Functional (FRDF) that represents the
energy of inhomogeneous $^4$He. The form of  $E_0 [\rho]$  and the discussion of
 formalism  can be found in Ref. \cite{gatica}. Such density functionals have previously been applied to the  study of 
adsorbed films \cite{films} and impurity atoms in liquid He. \cite{impu}

At zero temperature, the density of the $N$ helium atoms is $\rho (r) = N 
|\Psi(r)|^2$, where $\Psi(r)$ is the single particle ground state wave function
 that  minimizes the energy $E [\rho]$. The minimization procedure leads to 
the following Hartree-Fock equation for $\Psi$,
\beq
(-(\hbar^2/2m)\ \nabla^2+ U\ [\rho])\ \Psi= \mu\ \Psi
\label{eq:sch}
\eeq
Here $U [\rho]$ is the effective single particle interaction and is equal to
\beq
U [\rho] = \delta E/\delta \rho.
\eeq
The ground state wave function, $\Psi$,   depends only on the distance
to the axis of the nanotube, $r$, and equation \ref{eq:sch} becomes
 one-dimensional. We compute $\Psi$ and $\mu$ by solving equation \ref{eq:sch}
 self-consistently. The  results are displayed in figures \ref{fig:DF}-\ref{fig:energy}
, for the 
6 \AA\ nanotube.

In figure \ref{fig:DF} we plot the density as a function of the radial
coordinate $r$ for different values of the chemical potential. For small $\mu$ the
 density  has only one peak centered at $R=3.06$ \AA. As long as
the value of $\mu$ increases, the height of this peak grows,
 reaching densities  much larger than the saturation density
 of bulk $^4$He (0.022 \AA$^{-3}$), suggesting the formation of a
solid phase (which is not accurately described by our method).
   For larger $\mu$ the axial peak
becomes appreciable and continues to grow as $\mu$
 increases, while the shell peak remains constant.
 The picture agrees very well with the model we proposed
in section \ref{sec:hartree}.

We evaluated the number of atoms in the axial region,
$N_a$, and in the shell region, $N_s$, by integrating each
 of the peaks of the density. In figure \ref{fig:axial-shell_density} we 
plot the axial density $N_a/L$ and the shell density
$\theta=N_s/(L 2 \pi R)$ vs. $\mu$.
As we can see $N_a/L$ is non-zero for $\mu$ greater than
$\mu_c$ = -108 K.  This axial phase threshold
corresponds to $N/L$=2.2 \AA$^{-1}$ and to
 $\theta_c$=0.113 \AA$^{-2}$ in  agreement with our previous
Hartree model's 
prediction (see table \ref{table:4He}).
 
The energy per atom and the chemical potential $\mu$ are
displayed in figure \ref{fig:energy}.
Note a kink in the $\mu$ curve at the axial phase threshold. 
This effect can be
attributed to the presence of a ``gas-like'' axial phase, since
$d\mu/dN$ is proportional to the inverse of the compressibility.

\section{Conclusions}
\label{sec:conclusions}

In this work we have presented various calculations pertaining to a
hypothetical quasi-one-dimensional axial phase of quantum fluids adsorbed in 
nanotubes. Such a phase ought to be thermodynamically distinct from the high 
density cylindrical shell phase at lower density. Our calculations use an
approximate Hartree model for H$_2$ and the He isotopes. Low temperature path 
integral calculations for H$_2$ are qualitatively consistent with the Hartree 
results, apart from the case of very small R. For $^4$He, a density
functional model yields results which are also consistent with those of the
Hartree method. We find such an axial phase in all of these cases, except
for very small tubes containing H$_2$. We note that similar behavior is
expected for classical fluids consisting of small atoms, e.g. Ne and Ar. 
\cite{steele} Both our methodology and the results presented here are 
qualitatively similar to those obtained previously for the problem of 
determining the monolayer completion of quantum gases adsorbed on graphite 
\cite{whitlock,cheng}. Quantitatively, however, there is a big difference: the
threshold chemical potential values are lower for small tubes than for a
flat surface, due to the higher effective coordination of the particles in
small tubes.

We have not addressed here one of the most important questions: what are
the properties of this axial phase? We expect these to be entirely
different from those of the second layer on graphite because of the
difference in dimensionality. In particular, we anticipate the behavior to
be that of the so-called ``Luttinger liquid'' in one dimension. Among the
most remarkable of these properties for fermions are the absence of a fermi 
surface and the dominant role of spin and density fluctuations at low T. We 
hope that relevant calculations and experiments are soon forthcoming.

\acknowledgements{We are grateful to M. J. Bojan, M. Boninsegni, M. W. H. Chan,
V. H. Crespi, P. C. Eklund, R. B. Hallock, S. Hernandez, 
W. A. Steele and K. Williams for helpful discussions. This research has been supported by ARO,
 NSF, the University of Buenos Aires, and the Petroleum Research Fund of the 
American Chemical Society.}

\newpage
{\bf APPENDIX Curvature correction to energy}\\

We here compute the leading curvature correction to the potential energy
$P(R)$ per atom located on a cylindrical surface of infinite length and
radius $R$. It is typical in such cases to find an expansion of the form:
\beq
\frac{P(R)}{P(\infty)} \sim 1 + \left(\frac{b}{R}\right)^2
\eeq
where $b$ is a characteristic length of the system, assumed to be small
compared to $R$ in the present asymptotic limit. We shall confirm this
expectation here.

It is reasonable to identify the curvature correction as arising primarily
from the attractive part of the interatomic interaction since that
experiences a larger distance scale than the repulsive part of the
interaction and so becomes comparable to $R$ first as the graphene sheet is curved.
The attraction varies as
\beq
V(r) \sim -\frac{C}{r^6}
\eeq
where $C$ is the interatomic dispersion coefficient. For the case of atoms
having 2d density $\theta$,
\beq
-2 \frac{P(R)}{C \theta} = R \int dz \int d\varphi\ \frac{1}{[z^2 + (2 R
sin(\varphi/2)^2]^3}.
\eeq
where the 2 on the left side avoids double-counting and we bear in mind
that eventually we will need a small distance cutoff ($r=a$) in order to get
a finite answer. The planar limit, $R=\infty$, involves only small $\varphi$
contributions, leading to
\beq
-2 \frac{P(\infty)}{C \theta} = \int dz \int dx\ (z^2 + x^2)^{-3} = 
\frac{\pi}{2 a^4}
\eeq
The difference $\Delta P$ between this value and that of P at finite $R$ may be
evaluated with an asymptotic expansion. We find
\beq
-8 \frac{\Delta P}{C\ \theta} = \frac{3\ \pi}{ 8\ a^2}
\eeq
so that the ratio
\beq
\frac{\Delta P}{P} = \frac{3}{16} \left(\frac{a}{R}\right)^2
\eeq
confirming the anticipated curvature dependence. A plausible near-neighbor
cutoff is $a=(\pi/\theta)^{1/2} \sim 2$ \AA. For a nanotube of $R_C = 7$ \AA, 
the film is situated at $R \sim 4$ \AA. Then the curvature correction to the 
energy per atom is about 5\%.

\newpage
\begin{table}[h]
\caption{Properties of $^4$He atoms in the axial and shell states at threshold 
for formation of the axial phase,
 as a function of radius $R_C$ of the nanotube. The quantities $r_{rms}$, 
$V_C$, $V_{tot}$, $r_{min}$, $\mu_c$, $R$, and $E_0$ are defined in the 
text.}
\vspace*{0.2 in}
\begin{tabular} {cccc} 
$R_C$ (\AA)&                  6&       7&       8\\ \hline
$r_{rms}$ (\AA)&           0.41&    1.01&    2.00\\
-$V_{tot} (r_{min})$ (K)&   166&      97&      79\\
-$V_C (r_{min})$ (K)&        60&      40&      33\\
$r_{min}$ (\AA)   &        0.00&    1.25&    2.41\\
$\theta_c$ (\AA$^{-2}$) & 0.113&   0.114&   0.117\\
-$\mu_c$ (K)&                98&      75&      54\\
-$E_0$ (K)  &               197&     180&     170\\
$R$ (\AA)&                 3.06&    4.07&    5.07\\
\end{tabular}
\label{table:4He}
\end{table}

\begin{table}[h]
\caption{Properties of $^3$He atoms in the axial and shell states at 
threshold, as in Table \ref{table:4He}.}
\vspace*{0.2 in}
\begin{tabular} {cccc} 
$R_C$ (\AA)&                  6&       7&       8\\ \hline
$r_{rms}$ (\AA)&           0.56&    1.03&    1.96\\
-$V_{tot} (r_{min})$ (K)&   155&      95&      77\\
-$V_C (r_{min})$ (K)  &      60&      40&      32\\
$r_{min}$ (\AA)   &        0.00&    1.38&    2.41\\
$\theta_c$ (\AA$^{-2}$) & 0.110&   0.110&   0.113\\
-$\mu_c$ (K)&                88&      72&      51\\
$R$ (\AA)&                 3.04&    4.05&    5.02\\
\end{tabular}
\label{table:3He}
\end{table}

\newpage
\begin{table}[h]
\caption{Properties of H$_2$ molecules in the axial and shell states at 
threshold, as in Table \ref{table:4He}.}
\vspace*{0.2 in}
\begin{tabular} {cccc} 
$R_C$ (\AA)&                  6&       7&       8\\ \hline
$r_{rms}$ (\AA)&             --&    0.55&    1.38\\
-$V_{tot} (r_{min})$ (K)&  54.0&     437&     311\\
-$V_C (r_{min})$ (K)&       254&     141&     106\\
$r_{min}$ (\AA)&           0.00&    0.44&    1.67\\
$\theta_c$ (\AA$^{-2}$) &    --&   0.097&   0.099\\
 -$\mu_c$ (K)&               --&     385&     261\\
-$E_0$ (K)&                 749&     689&     651\\
$R$ (\AA)&                 2.69&    3.67&    4.65\\
\end{tabular}
\label{table:H2}
\end{table}

\begin{table}[h]
\caption{Properties of a $^3$He-$^4$He mixture in the axial and shell states 
at threshold, as in Table \ref{table:4He}.}
\vspace*{0.2 in}
\begin{tabular} {cccc} 
$R_C$ (\AA)&                  6&       7&       8\\ \hline
$r_{rms}$ (\AA)&           0.43&    1.00&    1.92\\
-$V_{tot} (r_{min})$ (K)&   157&      97&      79\\
-$V_C (r_{min})$ (K)&        60&      32&      19\\
$r_{min}$ (\AA)&           0.00&    1.36&    2.35\\
$\theta_{3,c}$ (\AA$^{-2}$)&0.110&  0.109&   0.112\\
$\theta_{4,c}$ (\AA$^{-2}$)&0.114&  0.114&   0.117\\
-$\mu_c$ (K)&                88&      72&      51\\
\end{tabular}
\label{table:mixture}
\end{table}

\newpage
\begin{figure}[ht]
\caption{Upwards (downwards) sloping curves are the chemical potentials of the 
cylindrical (axial) phases for $^4$He as a function of 2d shell density, 
$\theta_4$, in the cases of nanotubes with radii $R_C$ = 6, 7, 8 \AA, from 
bottom to top.}
\label{fig:mu4}
\end{figure}

\begin{figure}[ht]
\caption{Same as Fig. \ref{fig:mu4}, but for $^3$He.}
\label{fig:mu3}
\end{figure}

\begin{figure}[ht]
\caption{Hartree model calculations of probability density (unnormalized) for 
$^4$He atoms in tubes of radius of 6 \AA\ (---), 7 \AA\ (--- ---) and 
8 \AA\ (--- $\cdot$ ---).} 
\label{fig:rho}
\end{figure}

\begin{figure}[ht]
\caption{Curves are as in Fig. \ref{fig:mu4}, except that no curve exists for 
$R_C$ = 6 \AA, as discussed in text.}
\label{fig:mu_H2}
\end{figure}

\begin{figure}[ht]
\caption{Hartree model calculations, discussed in text, of probability 
density for H$_2$ in tubes of radius $R_C$= 6 \AA\ (---), 7 \AA\ (--- ---) 
and 8 \AA\ (--- $\cdot$ ---). Note the absence of the axial phase for the 
case of a 6 \AA\ tube.}
\label{fig:rho_H2}
\end{figure}

\begin{figure}[ht]
\caption{Shell density ($\theta$) and axial density ($N_a/L$) 
in a 7 \AA\ radius nanotube at a temperature 
of 10 K as computed by path integral grand canonical Monte Carlo.}
\label{fig:7Acoverage}
\end{figure}

\begin{figure}[ht]
\caption{Shell density ($\theta$) and axial density ($N_a/L$) 
in a 6 \AA\ radius nanotube at a temperature 
of 10 K as computed by path integral grand canonical Monte Carlo.}
\label{fig:6Acoverage}
\end{figure}

\begin{figure}[ht]
\caption{Density profile in reduced units for hydrogen in a 
6 \AA\ radius nanotube at 10 K. Note that the shell phase is 
progressively pushed outward as the axial phase is populated.}
\label{fig:6Arho}
\end{figure}

\begin{figure}[ht]
\caption{Density profile in reduced units for hydrogen in a 
7 \AA\ radius nanotube at 10 K.}
\label{fig:7Arho}
\end{figure}

\begin{figure}[ht]
\caption{Shell density ($\theta$) and axial density ($N_a/L$) 
in a 8 \AA\ radius nanotube at a temperature 
of 10 K as computed by path integral grand canonical Monte Carlo.}
\label{fig:8Acoverage}
\end{figure}

\begin{figure}[ht]
\caption{Density profile in reduced units for hydrogen in a 
8 \AA\ radius nanotube at 10 K.}
\label{fig:8Arho}
\end{figure}

\begin{figure}[ht]
\caption{Equilibrium densities of the coexisting $^3$He and $^4$He phases.}
\label{fig:dens34}
\end{figure}

\begin{figure}[ht]
\caption{Schematic picture showing the separation of He isotopes and 
$^3$He atoms ($\circ$) moving in the $^4$He ($\bullet$) regions.}
\label{fig:mixture}
\end{figure}

\begin{figure}[ht]
\caption{The chemical potential of $^3$He, $\mu_3$, as a function of coverage, 
 $\theta_4$, in the case when the axial $^3$He atoms go within the $^4$He
shell.}
\label{fig:mu34}
\end{figure}

\begin{figure}[ht]
\caption{Density profile  of  $^4$He at $T=0$ in the case of a 
nanotube of radius 6 \AA\, from FRDF calculation.}
\label{fig:DF}
\end{figure}

\begin{figure}[ht]
\caption{Axial phase total number density and  shell density 
 for $^4$He in the 6 \AA\ nanotube, from FRDF 
calculation.}
\label{fig:axial-shell_density}
\end{figure}

\begin{figure}[ht]
\caption{ Energy per particle and chemical potential for $^4$He in the 6\AA\ nanotube, from FRDF 
calculation. The dashed line shows
the value of N/L at which the axial phase appears.}
\label{fig:energy}
\end{figure}


\newpage
\begin{figure}[ht]
\epsfysize=5.in \epsfbox{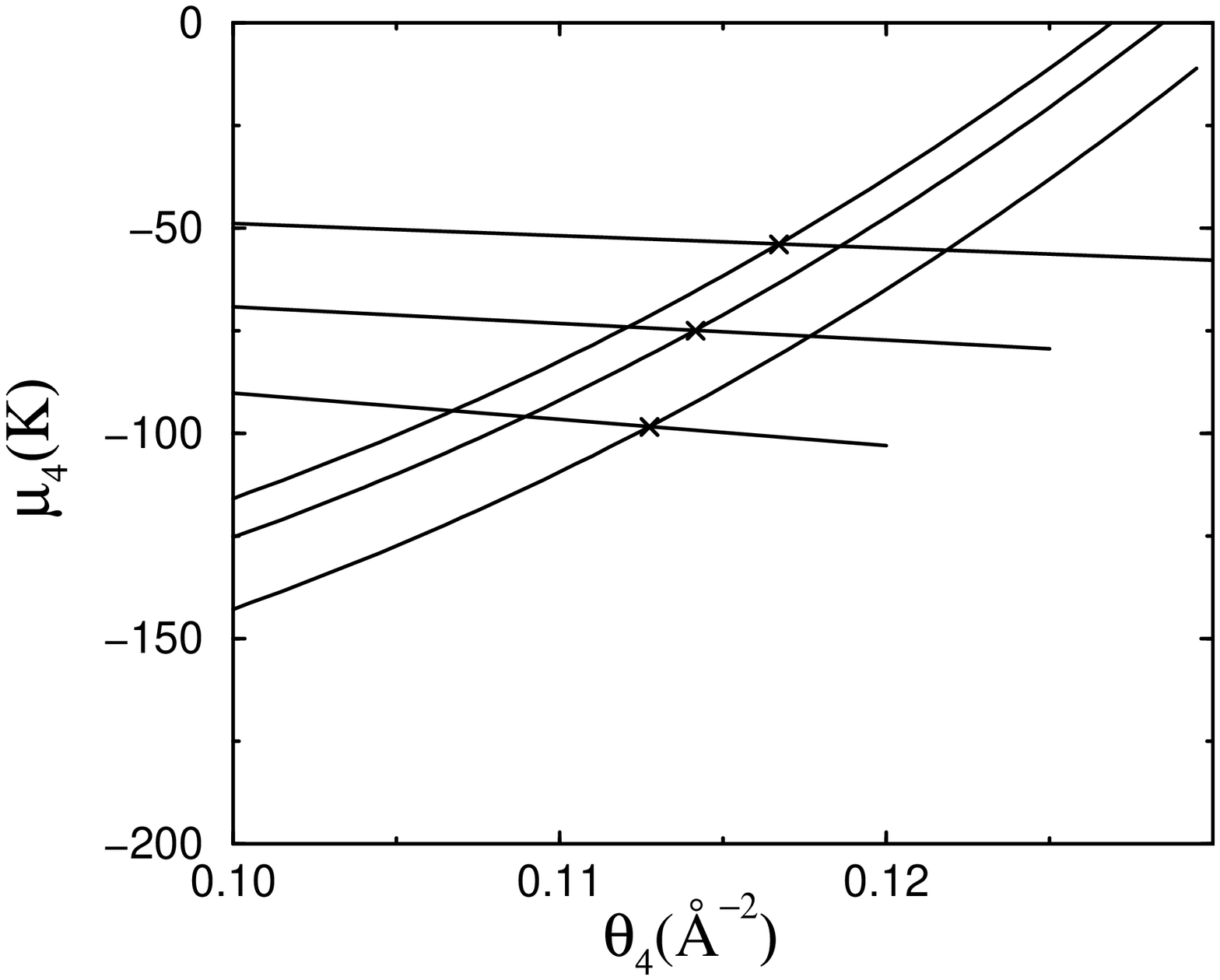}
\end{figure}
\vspace{5cm}
\begin{center}
{\bf FIG. 1}
\end{center}

\newpage
\begin{figure}[ht]
\epsfysize=5.in \epsfbox{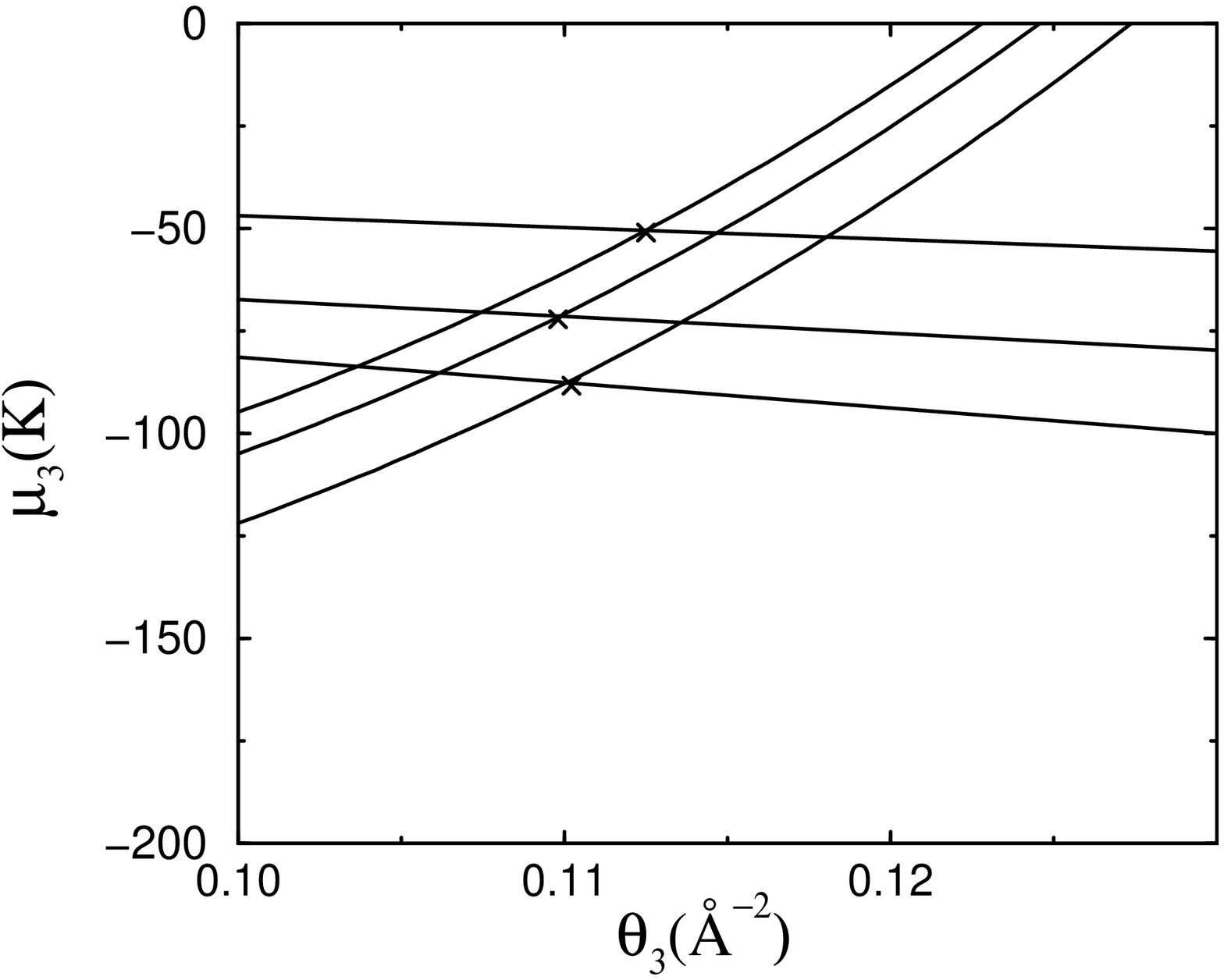}
\end{figure}
\vspace{5cm}
\begin{center}
{\bf FIG. 2}
\end{center}

\newpage
\begin{figure}[ht]
\epsfysize=5.in \epsfbox{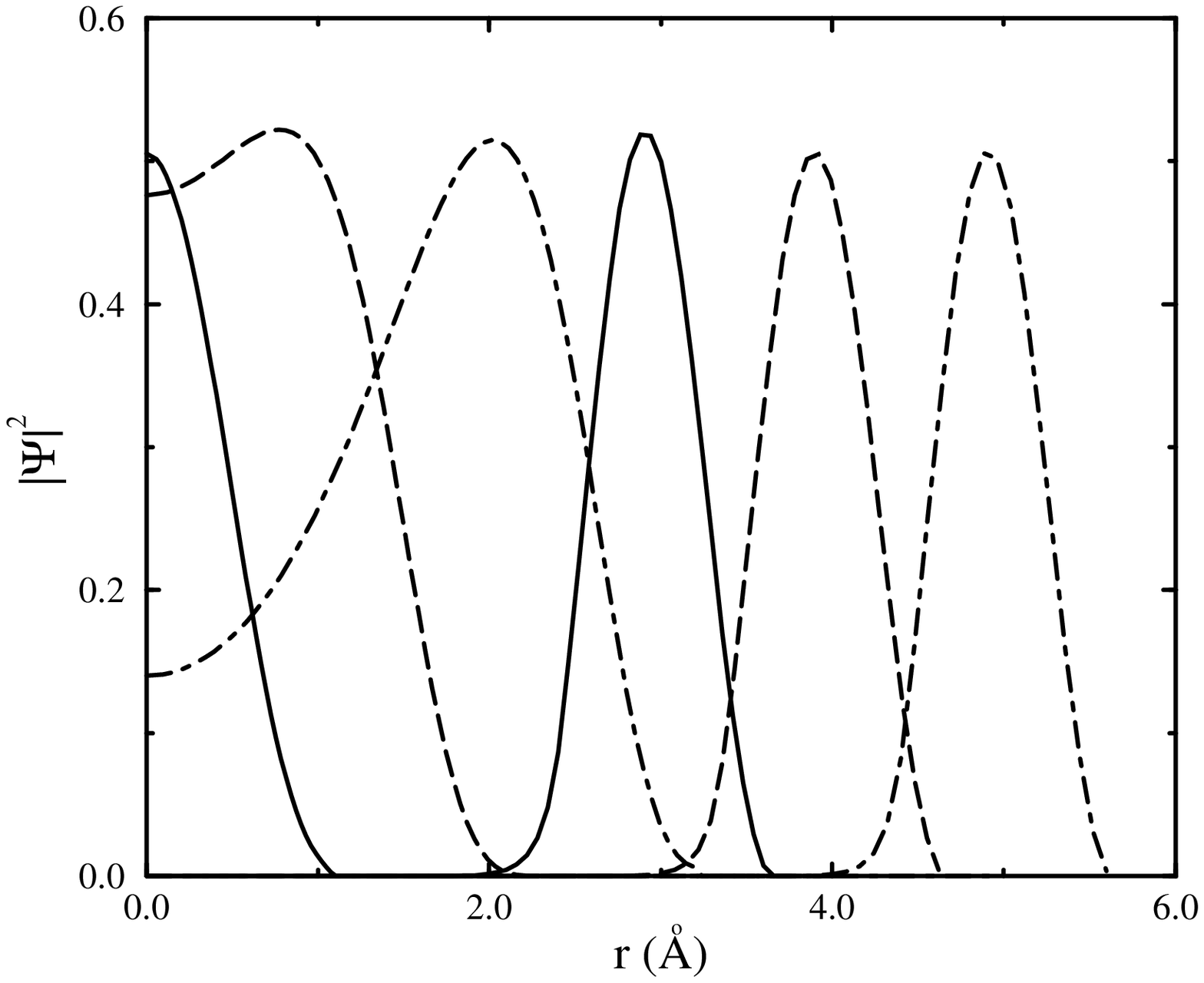}
\end{figure}
\vspace{5cm}
\begin{center}
{\bf FIG. 3}
\end{center}

\newpage
\begin{figure}[ht]
\epsfysize=5.in \epsfbox{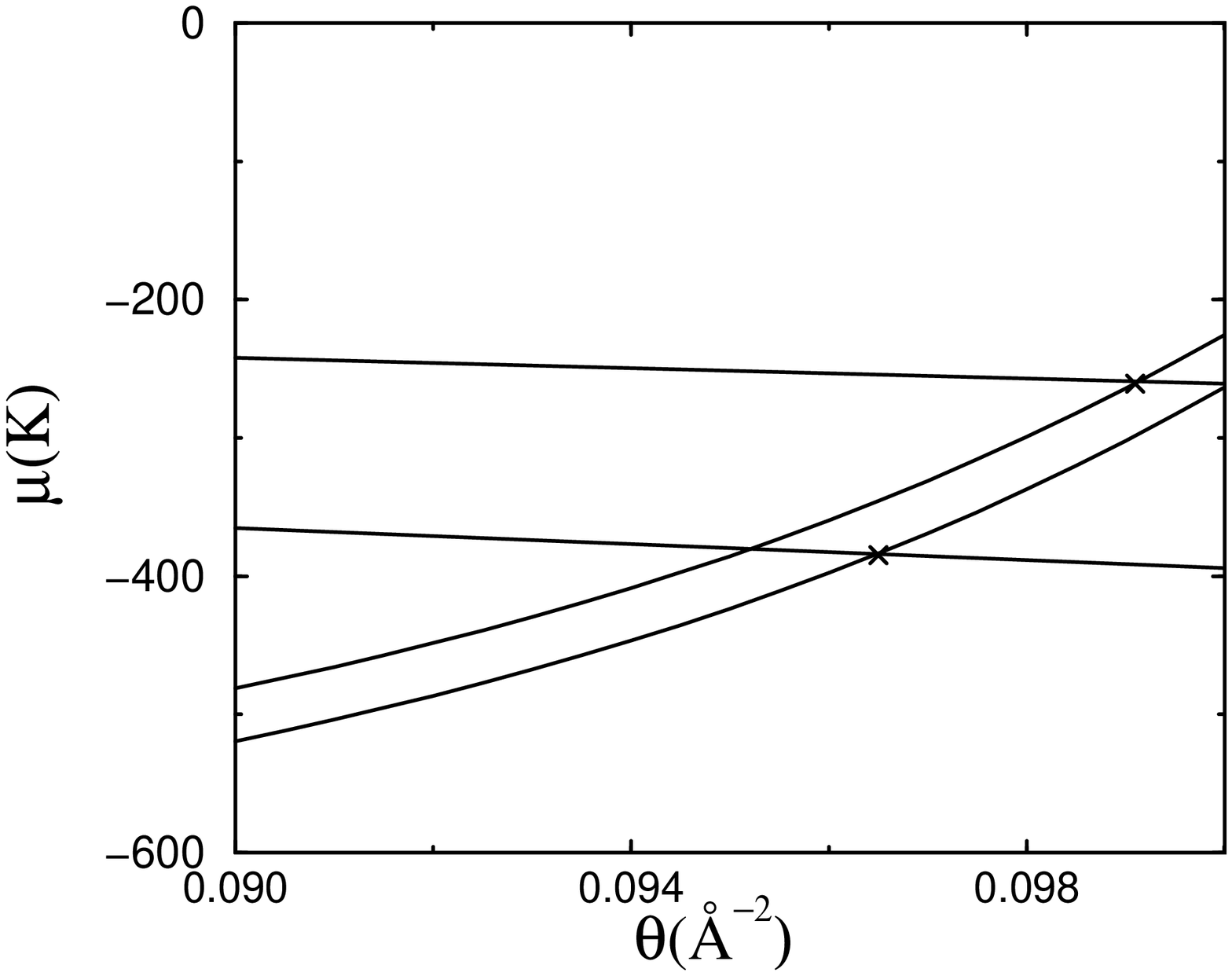}
\end{figure}
\vspace{5cm}
\begin{center}
{\bf FIG. 4}
\end{center}

\newpage
\begin{figure}[ht]
\epsfysize=5.in \epsfbox{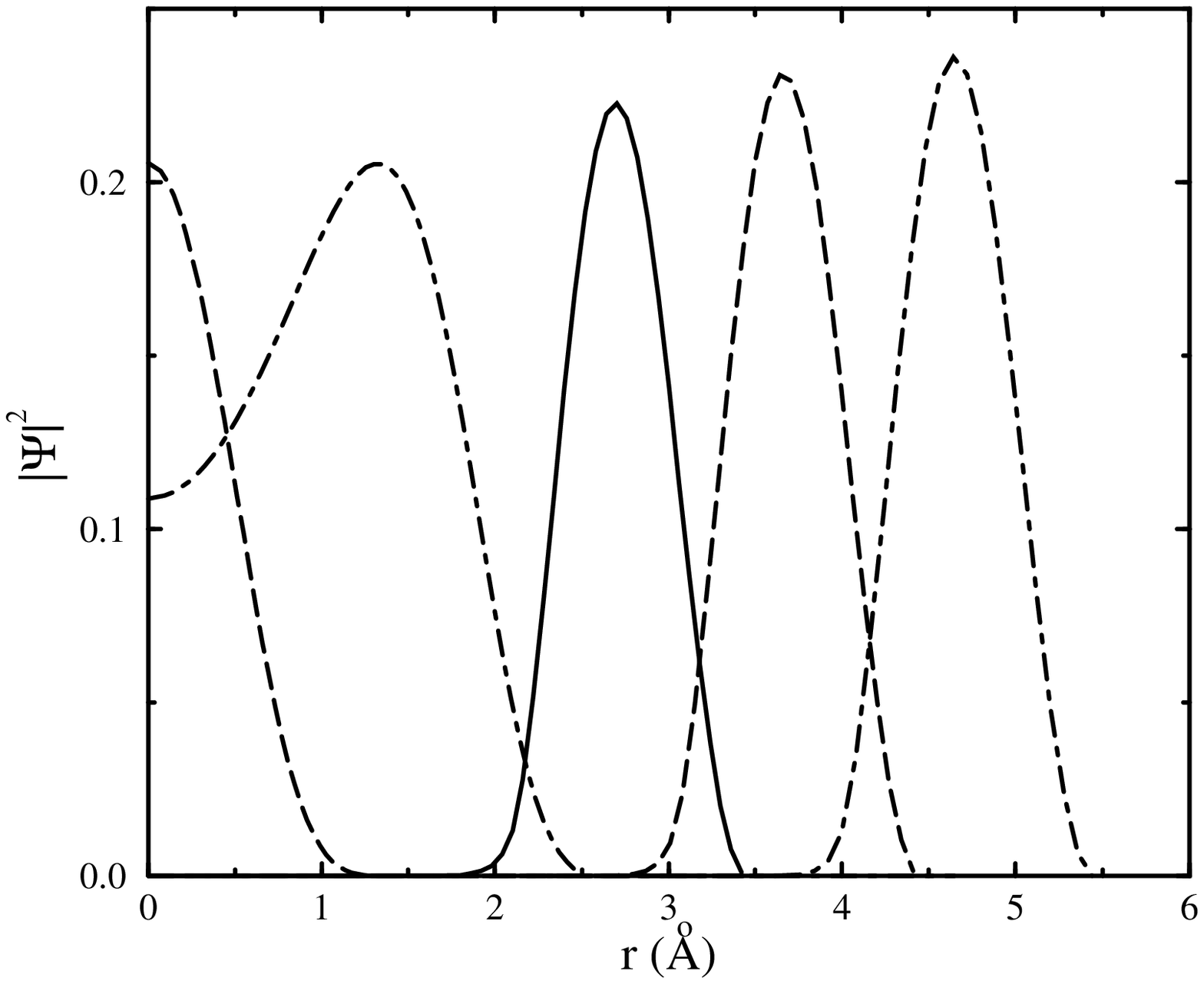}
\end{figure}
\vspace{5cm}
\begin{center}
{\bf FIG. 5}
\end{center}

\newpage
\begin{figure}[ht]
\epsfysize=5.in \epsfbox{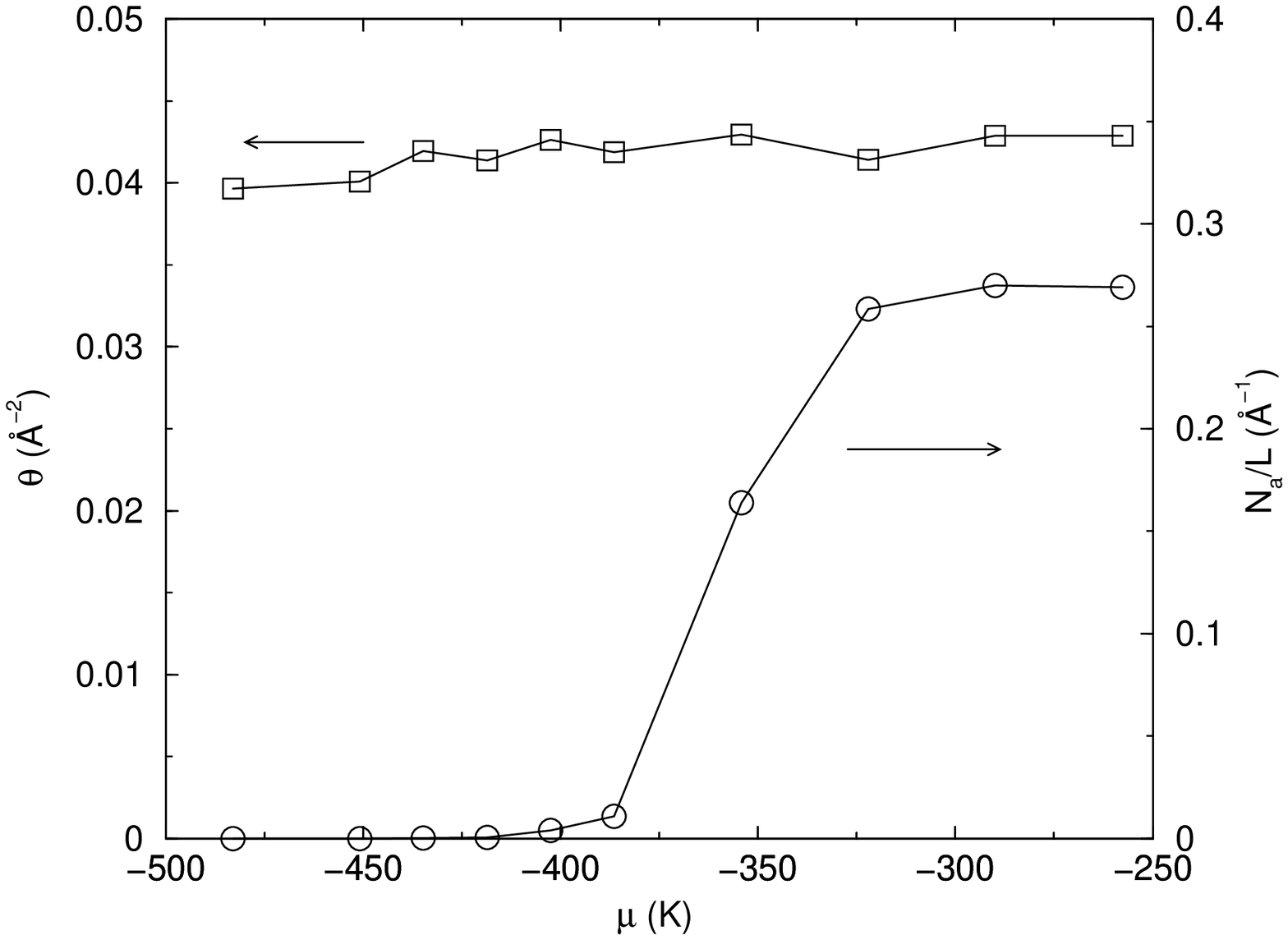}
\end{figure}
\vspace{5cm}
\begin{center}
{\bf FIG. \protect\ref{fig:7Acoverage}}
\end{center}

\newpage
\begin{figure}[ht]
\epsfysize=5.in \epsfbox{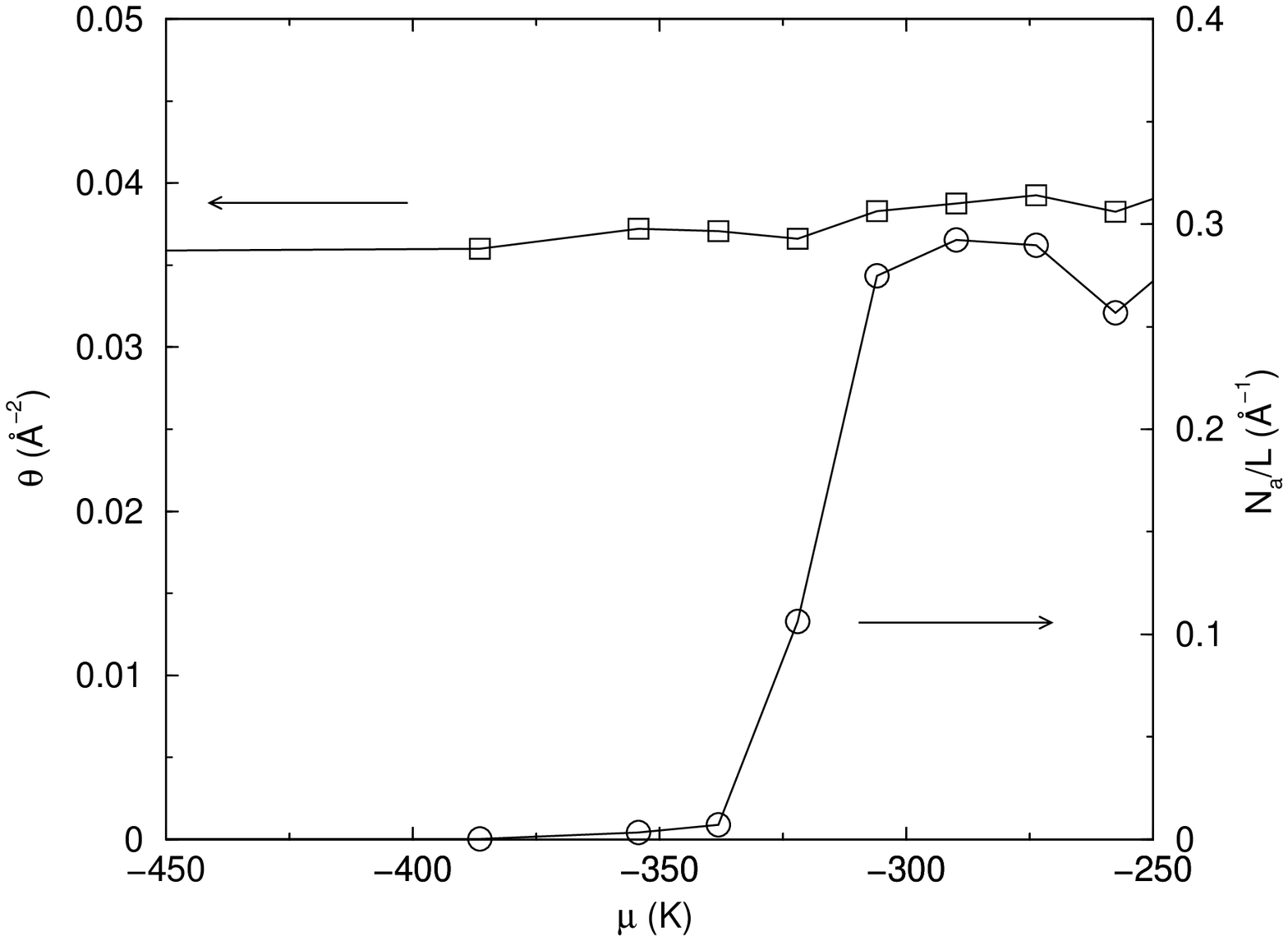}
\end{figure}
\vspace{5cm}
\begin{center}
{\bf FIG. \protect\ref{fig:6Acoverage}}
\end{center}

\newpage
\begin{figure}[ht]
\epsfysize=5.in \epsfbox{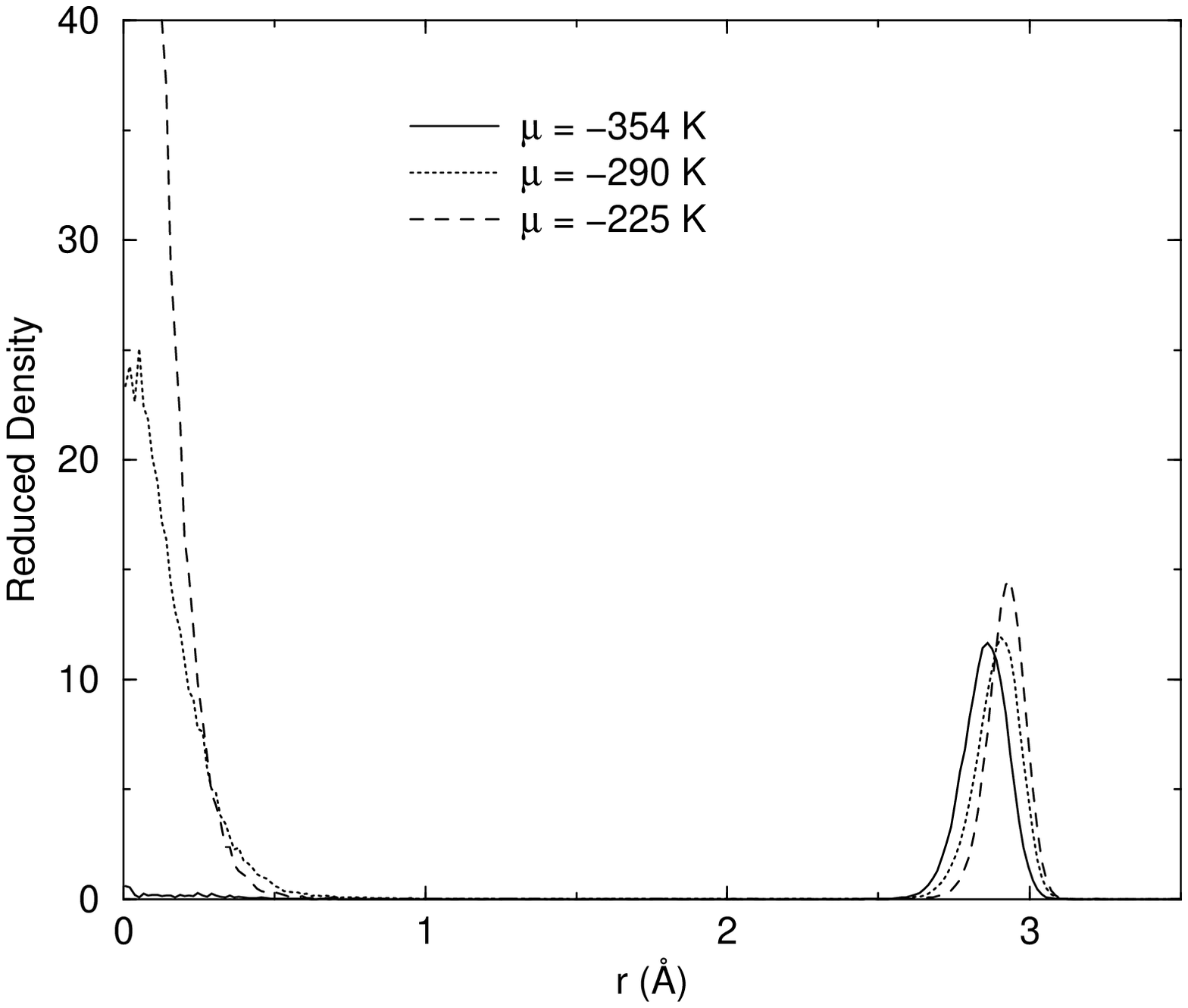}
\end{figure}
\vspace{5cm}
\begin{center}
{\bf FIG. \protect\ref{fig:6Arho}}
\end{center}

\newpage
\begin{figure}[ht]
\epsfysize=5.in \epsfbox{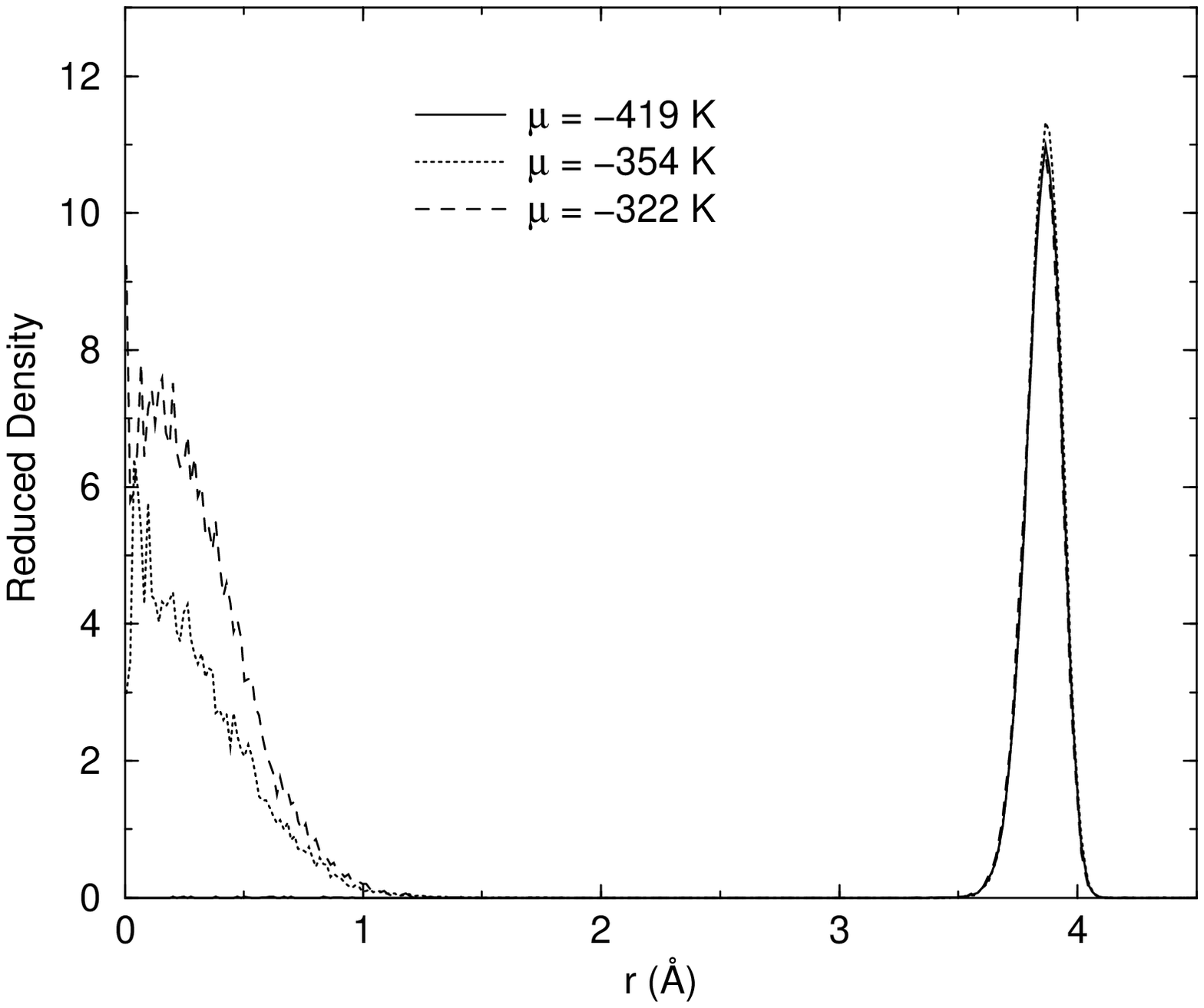}
\end{figure}
\vspace{5cm}
\begin{center}
{\bf FIG. \protect\ref{fig:7Arho}}
\end{center}

\newpage
\begin{figure}[ht]
\epsfysize=5.in \epsfbox{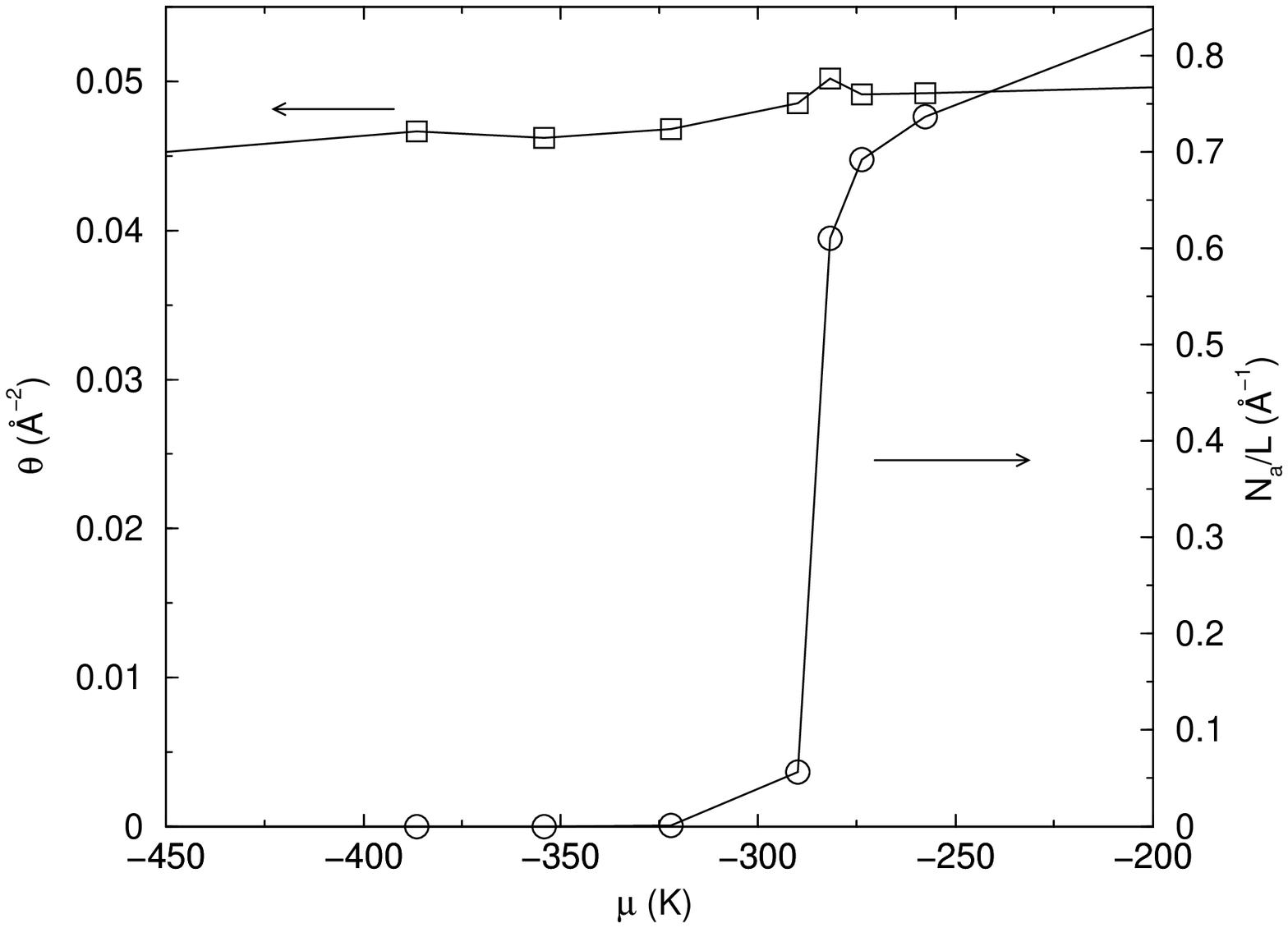}
\end{figure}
\vspace{5cm}
\begin{center}
{\bf FIG. \protect\ref{fig:8Acoverage}}
\end{center}

\newpage
\begin{figure}[ht]
\epsfysize=5.in \epsfbox{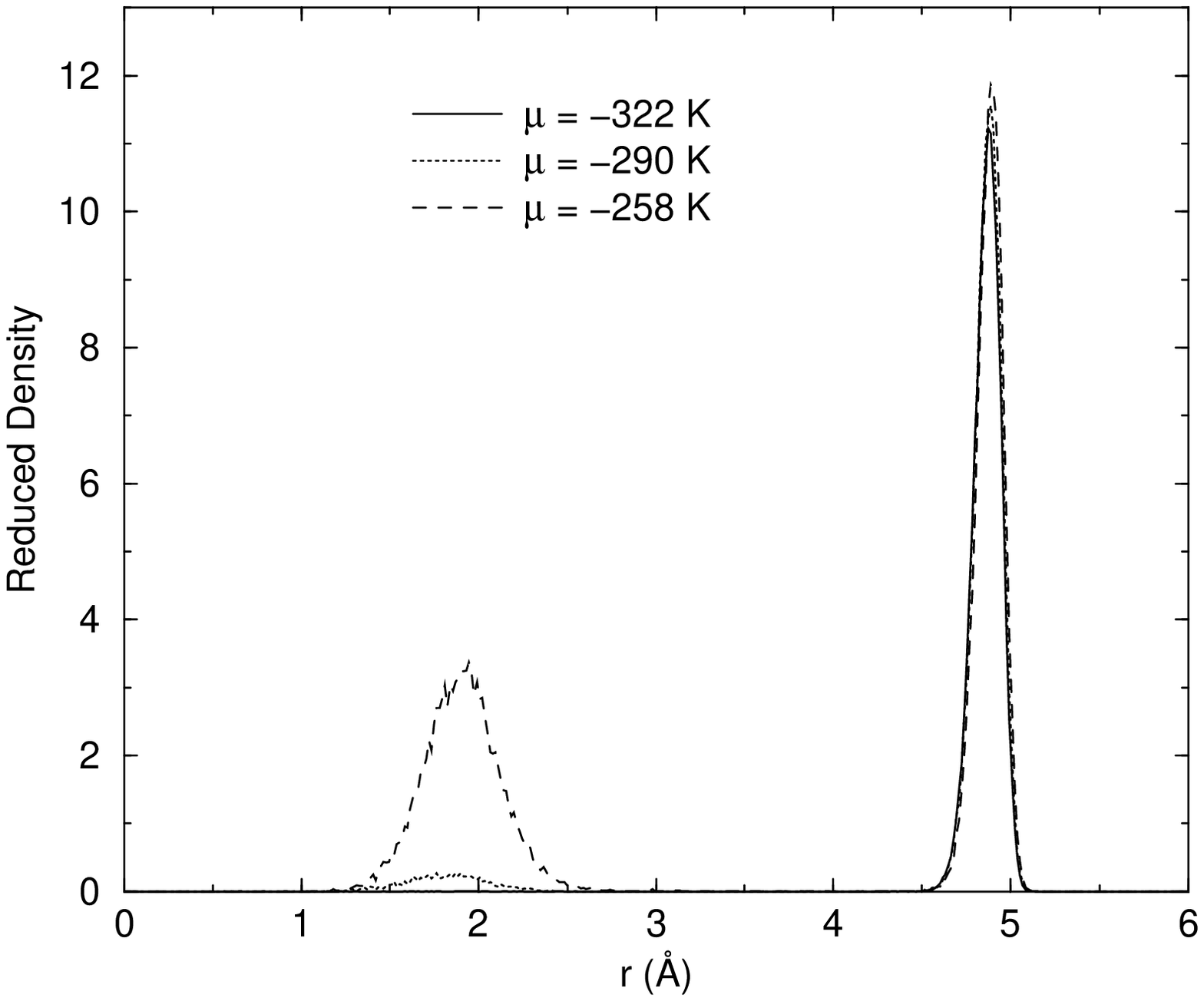}
\end{figure}
\vspace{5cm}
\begin{center}
{\bf FIG. \protect\ref{fig:8Arho}}
\end{center}

\newpage
\begin{figure}[ht]
\epsfysize=5.in \epsfbox{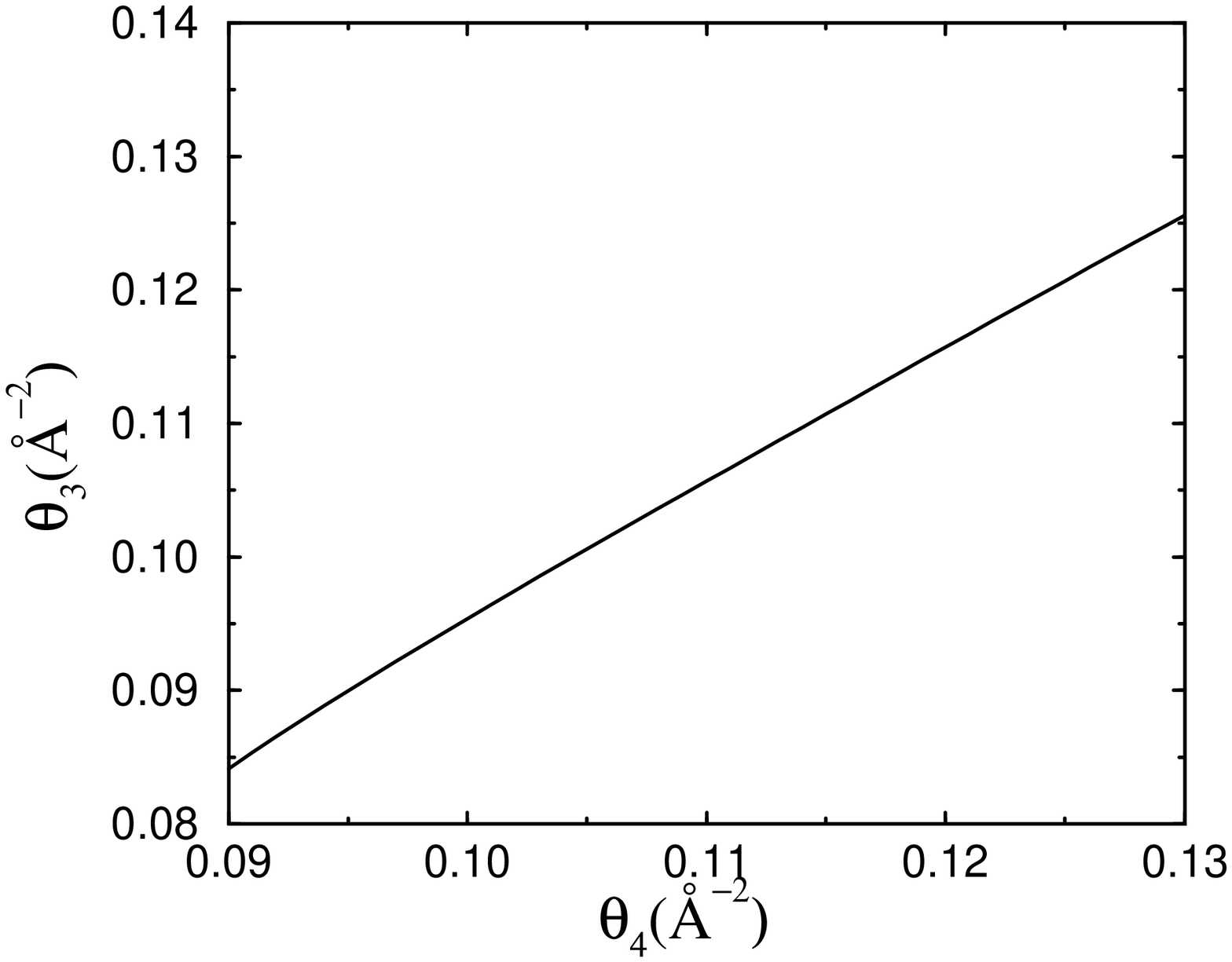}
\end{figure}
\vspace{5cm}
\begin{center}
{\bf FIG. \protect\ref{fig:dens34}}
\end{center}

\newpage
\begin{figure}[ht]
\epsfysize=1.5in \epsfbox{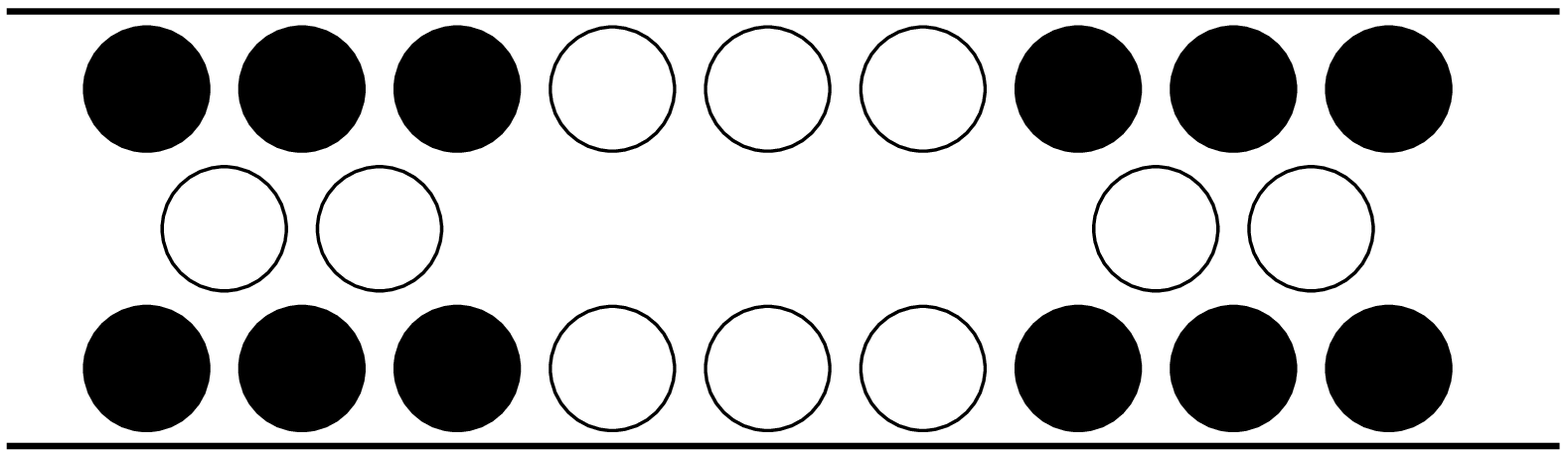}
\end{figure}
\vspace{5cm}
\begin{center}
{\bf FIG. \protect\ref{fig:mixture}}
\end{center}

\newpage
\begin{figure}[ht]
\epsfysize=5.in \epsfbox{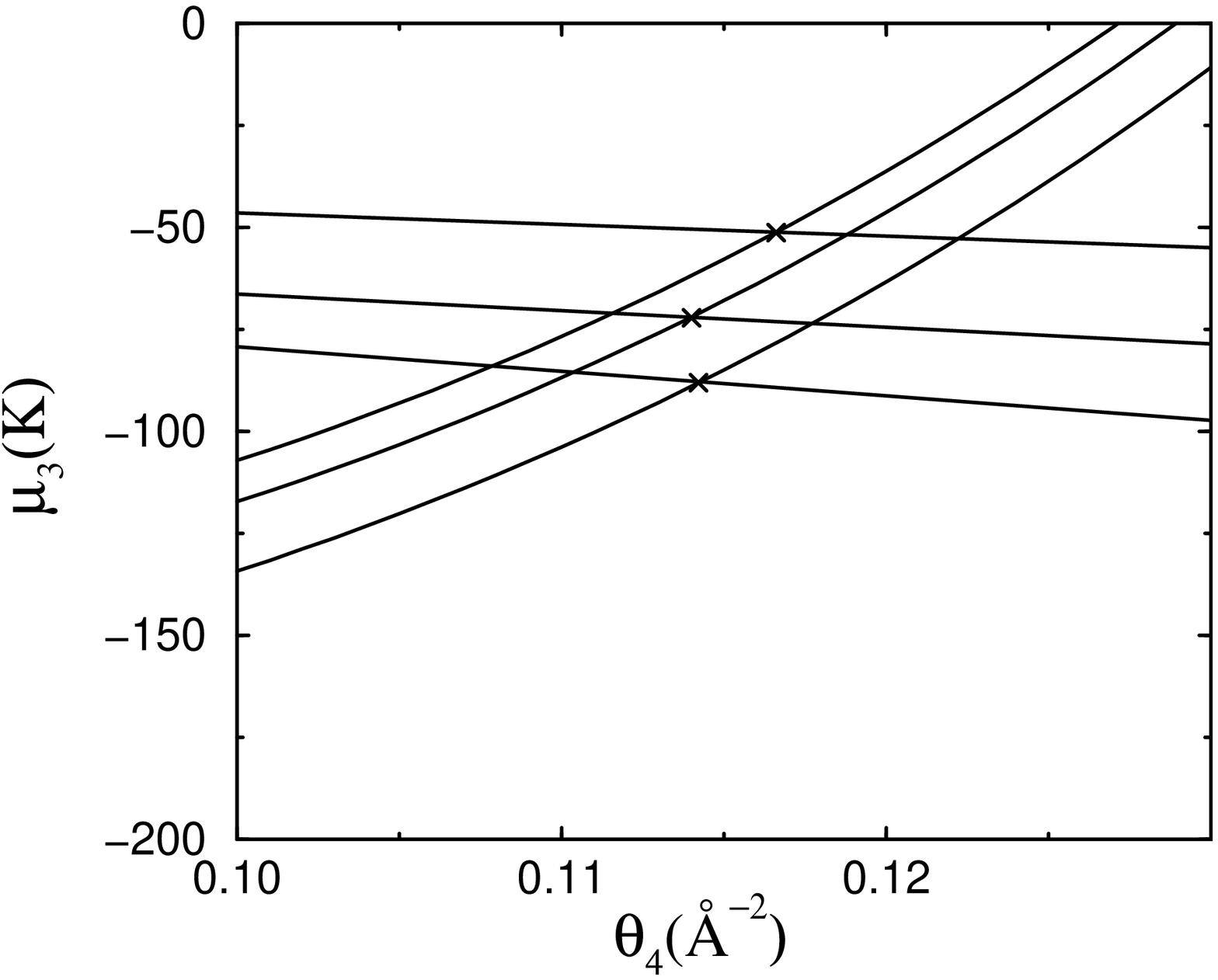}
\end{figure}
\vspace{5cm}
\begin{center}
{\bf FIG. \protect\ref{fig:mu34}}
\end{center}

\newpage
\begin{figure}[ht]
\epsfysize=5.in \epsfbox{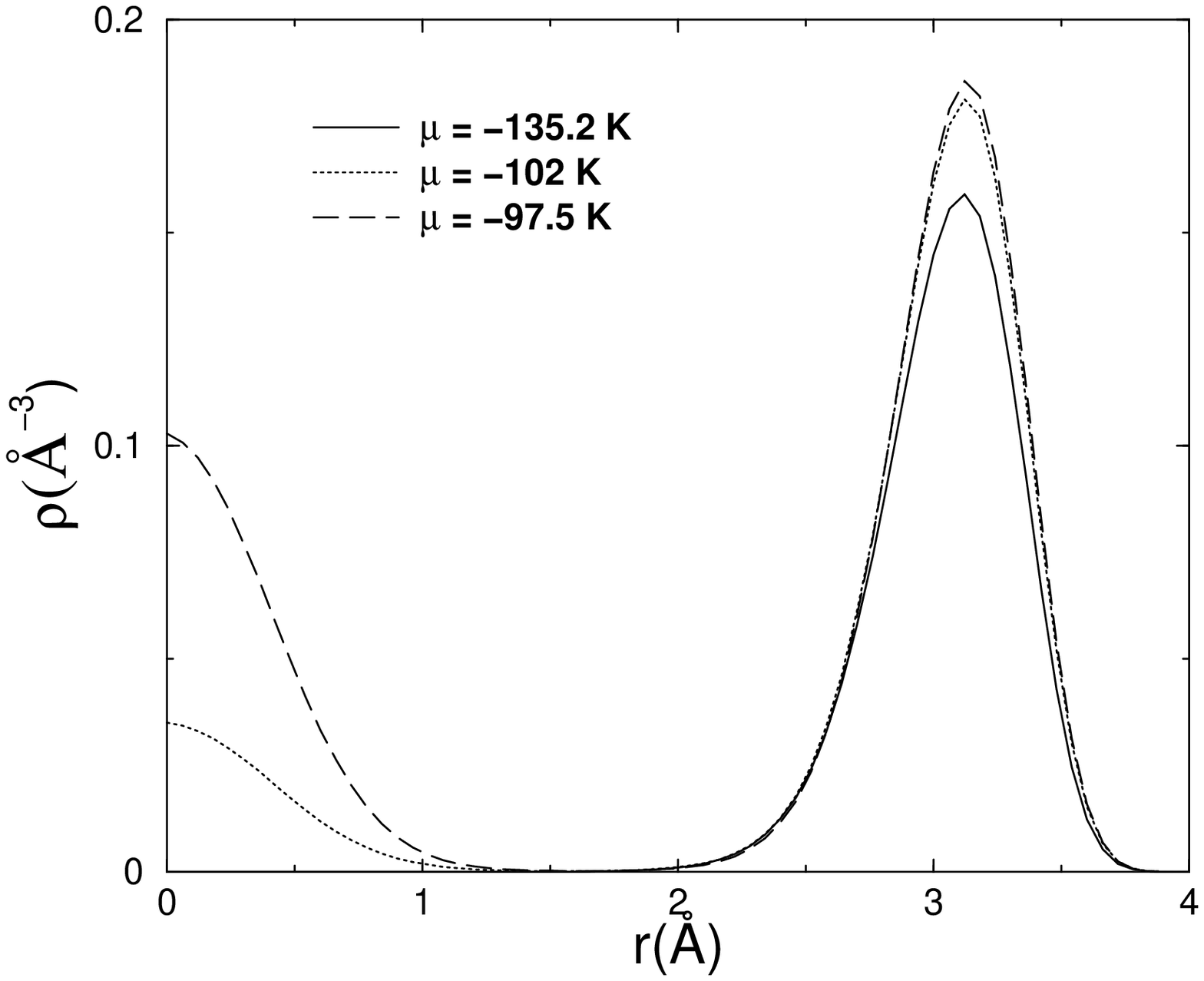}
\end{figure}
\vspace{5cm}
\begin{center}
{\bf FIG. \protect\ref{fig:DF}}
\end{center}

\newpage
\begin{figure}[ht]
\epsfysize=5.in \epsfbox{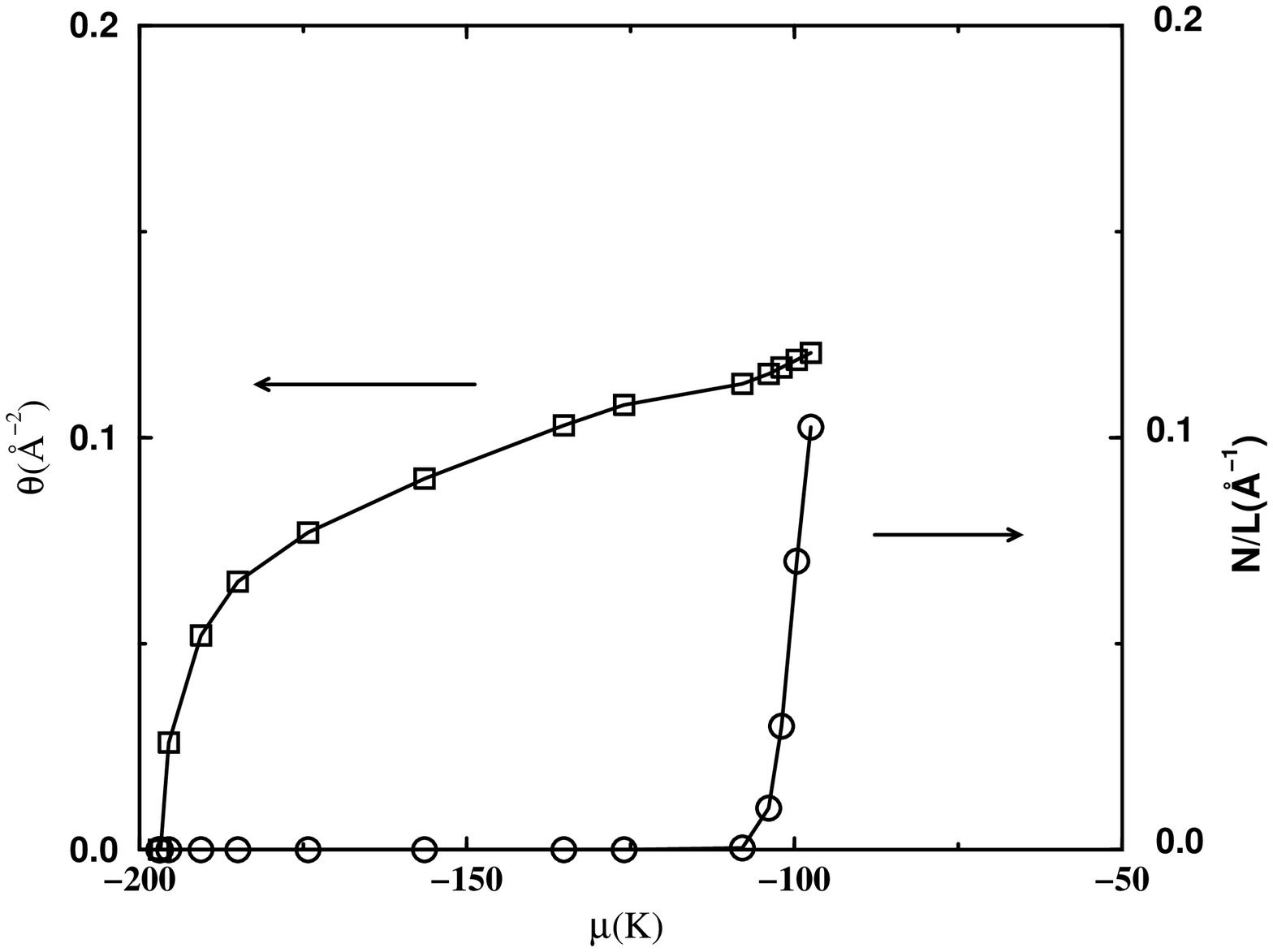}
\end{figure}
\vspace{5cm}
\begin{center}
{\bf FIG. \protect\ref{fig:axial-shell_density}}
\end{center}

\newpage
\begin{figure}[ht]
\epsfysize=5.in \epsfbox{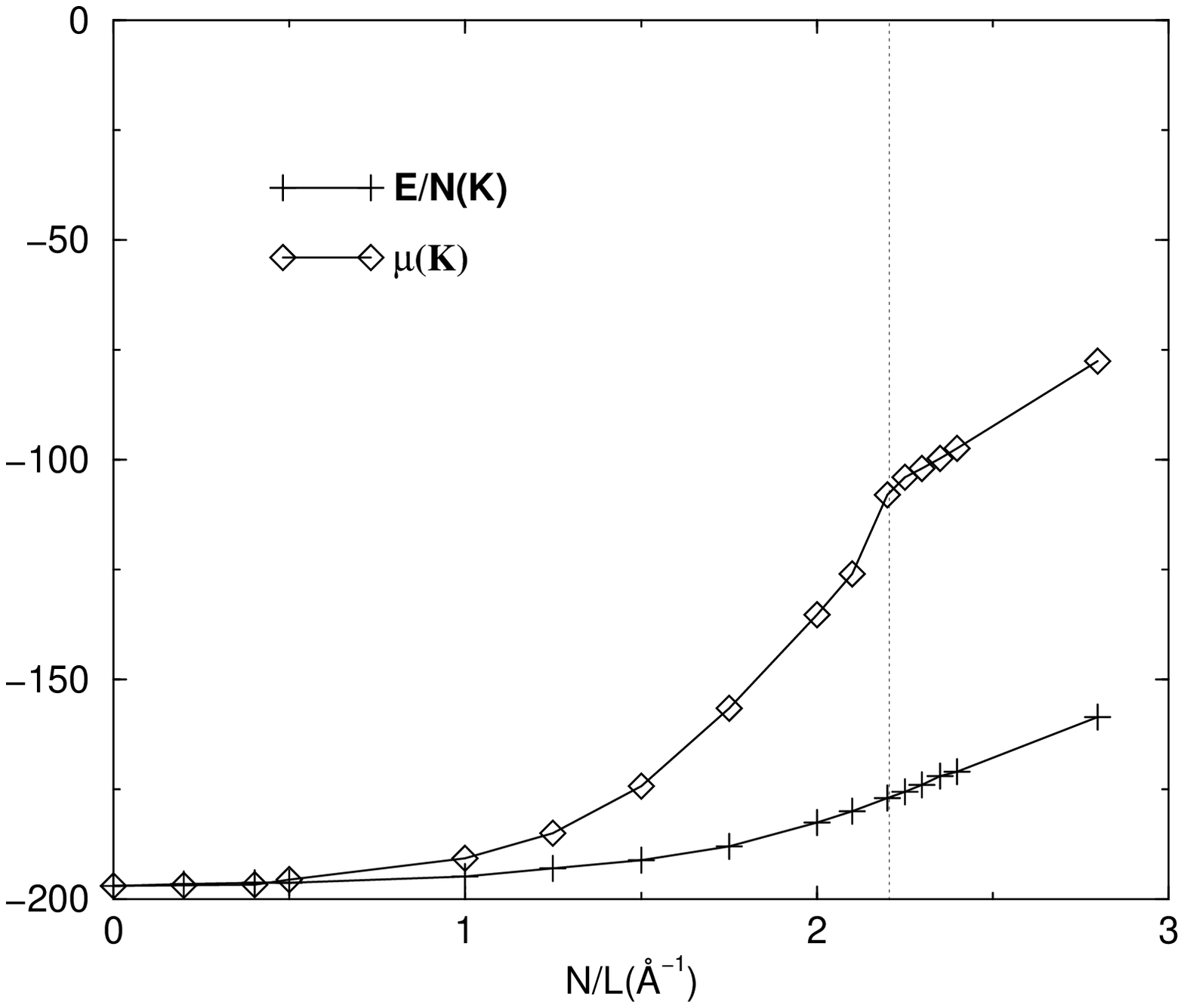}
\end{figure}
\vspace{5cm}
\begin{center}
 {\bf FIG. \protect\ref{fig:energy}}
\end{center}


\begin{thebibliography}{50}
\bibitem{ideal_gas}
An ideal gas regime at $T \equiv 0$ for He may or may not exist in a nanotube. 
Based on relevant points in the discussion below, we suspect that it does 
exist for $^4$He, but does not for $^3$He.
\bibitem{monolayer}
The corresponding values for $^3$He  on graphite are 0.108 \AA$^{-2}$ and
-27 K; see Refs. \cite{ni,bretz}.
\end{thebibliography}
\end{document}